\documentclass[twocolumn]{jpsj2} 
\title{Conserving Diagrammatic Approximations for  
Quantum Impurity Models:\\ NCA and CTMA}

\author{\textsc{J. Kroha}$^{1}$
\thanks{E-mail Address: kroha@th.physik.uni-bonn.de} and
\textsc{P. W\"olfle}$^{2}$
\thanks{E-mail address: woelfle@tkm.physik.uni-karlsruhe.de}} 

\inst{$^{1}$Physikalisches Institut, 
Universit\"at Bonn, 53115 Bonn, Germany\\
$^{2}$Institut f\"ur Theorie der Kondensierten Materie, Universit\"at
Karlsruhe, 76128 Karlsruhe, Germany}

\abst{Self-consistent diagrammatic approximations to the Anderson or
Kondo impurity model, using an exact pseudoparticle representation of the
impurity states, are reviewed. 
We first discuss the infrared exponents of the pseudoparticle 
propagators as indicators of Fermi liquid behavior through their 
dependence on the impurity occupation and on magnetic field.
Then we discuss the Non-Crossing Approximation (NCA), identifying its 
strengths, but also its fundamental shortcomings. 
Physical arguments as well as a perturbative renormalization
group analysis suggest that an infinite parquet-type
resummation of two-particle vertex diagrams, the Conserving 
T-Matrix Approximation (CTMA) will cure the deficiencies of NCA.
We review results on the pseudoparticle spectral functions, the spin
susceptibility and the impurity electron spectral function, 
supporting that the CTMA provides qualitatively correct results,
both in the high-temperature regime
and in the strong coupling Fermi liquid regime 
at low temperatures.
}

\kword{Kondo effect, Anderson impurity model, 
conserving approximations, Non-Crossing Approximation,
Conserving T-Matrix Approximation,
perturbative renormalization group}

\begin{document}
\maketitle

\section{Introduction}
\label{1}

Over the last 40 years the Kondo problem has become an archetypical
model of correlated electrons.  The discovery by Kondo
\cite{Kondo6469} of logarithmic anomalies in the perturbation series
for the electrical resistivity calculated within the spin exchange
model of a magnetic impurity in a metal gave but the first indication
of the existence of complex many-body effects in this and similar
models.  The key ingredient of these models is the availability of
local degrees of freedom at the impurity, e.g. the substates of a
local spin.  Coupling of the quantum impurity to the conduction
electrons of the metallic host induces a local interaction between the
conduction electrons, and generates delicate many-body resonance
states at low energy. 

Soon after Kondo's discovery it became clear that perturbation theory
in the antiferromagnetic exchange coupling constant $J$ is sufficient
only at high excitation energies/temperatures $T$ so that $JN_0\ell
n(D/T) \ll 1$, where $N_0 = \frac{1}{2D}$ is the conduction electron
density of states at the Fermi level and $D$ the half bandwidth
(high-energy cutoff).  By summing up the leading
logarithmic terms to all orders of perturbation theory, using the
perturbative renormalization group (RG) approach \cite{Anderson70}
it is found that a single dynamically generated energy scale, the
Kondo temperature $T_K \approx D\exp - \frac{1}{2N_0J}$ is generated.
The perturbative $RG$ holds as long as 
${\rm ln} (T/T_K) \stackrel{>}{\sim}  1$.

In the opposite limit of $T \ll T_K$, the impurity spin is found to be
screened by the conduction electron spins, and in the case of impurity
spin $S = \frac{1}{2}$ coupled to a single band of conduction
electrons, a spin singlet resonance state is formed.  This has been
established first beyond doubt within the numerical renormalization
group method (NRG) pioneered by K.G. Wilson \cite{Wilson75}.  Later an
analytical solution of the energy spectrum of the Kondo model was
obtained with the aid of the Bethe ansatz (BA) \cite{Andrei80,Wiegmann80},
fully confirming the NRG results.

While these highly successful exact solution methods have led to a 
virtually complete understanding
of the single-impurity Kondo problem, it became clear in the course of the
mid 1980s and 1990s that the Kondoesque enhancement of electron 
scattering at low energies plays a central role even in more complex
situations, like in strongly correlated lattice electron systems as well as
in mesoscopic devices. In the former systems, this is because the strong
on-site repulsion in conjunction with a very short correlation length
effectively induces local moment physics, as has been shown formally by the 
mapping of correlated lattice models onto quantum impurity 
models with a self-consistency condition by means of the 
Dynamical Mean Field Theory (DMFT) \cite{Georges96}.  In the latter systems,
discrete, localized quantum degrees of freedom, coupled to a continuum of
conduction electrons, are often formed because of the
spatial confinement of interacting electron states, e.g. 
in quantum dots \cite{Goldhaber98} and carbon nanotubes \cite{Nygaard01},
nanoscale constrictions and charge traps \cite{Ralph89}, 
or magnetic molecules \cite{Park02},    
leading to Kondo behavior in the electronic transport. 
All these findings have demonstrated that the Kondo effect is
an ubiquitous phenomenon in interacting electron systems and have made it
a central theme of condensed matter physics. 

To tackle the complex quantum impurity problems arising in such systems, 
it is desirable to develop, besides the
exact solution methods, flexible, approximate, but systematic
techniques which do not rely on special symmetry conditions or 
a relatively simple model structure, like NRG, or
on integrability conditions, like the BA, and which are still
capable of describing the high energy and low energy sectors of the
model as well as the crossover region around $T_K$. In many respects,
advanced perturbation theory methods are attractive here because of
their flexibility. They do not require a
simple conduction electron density of states (CDOS), and thus may be
employed as impurity solvers for the self-consistent quantum impurity
models of DMFT\cite{Georges96}, where the
CDOS may acquire considerable structure, e.g. a Mott-Hubbard gap and a
narrow resonance in the gap.  They may be applied to impurity models
with more complex structure, e.g. Anderson models with several levels
and realistic electron-electron interactions.  They may even be used
to describe quantum frustrated systems like multi-channel Kondo or
Anderson models. An additional advantage of a method based on
resummation of perturbation theory is that it may be easily
generalized to nonequilibrium systems \cite{Kroha02,Langreth03}.  
The latter is of interest in
the context of electron transport through nanostructures like quantum
dots or point contacts showing the Kondo effect.

The perturbation theory for Kondo or Anderson models is complicated
by the fact that the spin operators do not obey canonical commutation
rules or that the electron dynamics on the Anderson impurity are 
effectively restricted to single occupancy by the strong Coulomb
repulsion, respectively. 
This difficulty can be overcome efficiently by representing the
quantum impurity degrees of freedom by canonical auxiliary particle fields
\cite{Abrikosov65,Barnes76}, rendering Wick's theorem valid.   
For a description of the strongly correlated Fermi liquid (FL) ground state 
an infinite resummation of the logarithmic terms of perturbation theory
is clearly necessary, where, however, the correct selection of terms is
crucial. As will be seen, this can be achieved systematically by
means of conserving approximations, derived by functional derivative
from a generating functional.   

In the present article we review the method of conserving approximations
for Kondo-like models in terms of auxiliary particle fields, 
with special emphasis on describing, by means of a single approximation
scheme, the perturbative regime at high energies as well as the strong 
coupling regime below $T_K$.
We review the Non-Crossing Approximation (NCA) as the simplest of a
hierarchy of conserving approximations and discuss its strengths and
its fundamental failures. A detailed analysis will reveal the
origin of the latter. This will lead to the Conserving T-Matrix 
Approximation (CTMA), which will be demonstrated to 
remedy the shortcomings of the NCA both in the high-energy and in the
low-energy strong coupling regime.

\section{Model and renormalized perturbation theory for constrained dynamics}
\label{2}

In many situations the Kondo physics is more clearly described in
terms of the Anderson impurity model \cite{Anderson61},  
\begin{eqnarray}
\label{eq:hamiltonian}
H&=& \sum_{\vec{k} \sigma} \epsilon_{\vec{k}} 
c^{\dagger}_{\vec{k} \sigma}c^{ }_{\vec{k} \sigma}\,+\, 
\sum_{\sigma}(\epsilon_{d}-\sigma B/2)
d_{\sigma}^{\dagger}d_{\sigma}^{ }  \\
&+& V \sum_{
\vec k \sigma } (c^{\dagger}_{\vec k \sigma } d_{\sigma}^{ } 
+ h.c.) + U d_{\uparrow}^{\dagger}d_{\uparrow}^{ }
            d_{\downarrow}^{\dagger}d_{\downarrow}^{ } \nonumber 
\end{eqnarray}
where $c^{\dagger}_{\vec{k} \sigma}$ and $d^{\dagger}_{\sigma}$ are the
creation operators for electrons with spin $\frac{1}{2}$
(spin degree of freedom $\sigma=\pm 1$) in a conduction band
state $\vec k$ and in the impurity level $\epsilon_d$, respectively.
To be general, we have included a magnetic field ${\cal B}$ acting
on the impurity spin with Zeeman energy $B=g\mu_B{\cal B}$
($\mu _B$ and $g$ are the Bohr magneton and the Land\'e factor,
respectively).   
The electrons may hop from the conduction band onto and off the impurity
with amplitude $V$.  A sufficiently large Coulomb interaction $U$ at
the impurity essentially prevents double occupancy of the impurity.
Provided that the impurity level is sufficiently far below the Fermi
energy $E_F$, its occupation number will be close to one, meaning that
a spin $S = \frac{1}{2}$ is located at the impurity. The Kondo spin
exchange interaction model follows from the Anderson model in the
limit of nearly single electron occupancy, after projecting out the
high energy sector \cite{Schrieffer66}.  

Since $V/D \ll
1$ usually, it appears natural to employ perturbation theory in $V$.
This perturbation theory is complicated by two facts:  (a) the
impurity is an interacting electron system, for which the powerful
quantum field theoretical methods like Wick's theorem, Feynman
diagrams and renormalization of propagators and vertices are not
immediately available; (b) the perturbation theory in $V$ is
characterized by logarithmically diverging terms, like the ones in
the exchange coupling $J$. 

\subsection{Pseudoparticle representation}
Problem (a) may be circumvented by working with pseudoparticle
representations for the impurity states \cite{Barnes76} 
(or equivalently a resolvent operator formalism), 
\begin{eqnarray}
\label{eq:sbrepresentation}
d^{\dagger}_{\sigma}&=&f^{\dagger}_{\sigma}b+\sigma a^{\dagger}f_{-\sigma}\ ,
\end{eqnarray}
where $f^{\dagger}_{\sigma}$ is the fermionic creation operator for
the singly occupied impurity state with spin $\sigma$ and
$b^{\dagger}$, $a^{\dagger}$ are the bosonic creation operators for  
the empty and doubly occupied impurity state, respectively.
Since the pseudoparticle representation necessarily
enlarges the Hilbert space into unphysical regions, care has to be
taken to project onto the physical subspace, defined by all many-body states
with pseudoparticle number 
$Q=\sum _{\sigma} f^{\dagger}_{\sigma}f^{ }_{\sigma} + 
b^{\dagger}b^{ } + a^{\dagger}a^{ }=1$.
This can be done in an elegant way for any expectation value
of physical operators acting on the impurity states by 
working in the grand canonical ensemble with respect to the conserved
charge $Q$ and simply taking the negative
chemical potential $\lambda$ of the pseudoparticles to infinity
\cite{Abrikosov65,Barnes76} at the end of the calculation, 
e.g. for the impurity or $d$-electron Green's function in the imaginary 
time domain,
\begin{eqnarray}
\label{eq:projection}
G_{d\sigma}(\tau)=- \lim _{\lambda\to\infty} 
\frac{\langle d_{\sigma}^{ }(\tau)d_{\sigma}^{\dagger}(0)
e ^{-\beta[H+\lambda(Q-1)]}\rangle}
{\langle Q e ^{-\beta[H+\lambda(Q-1)]} \rangle} \ .
\end{eqnarray}
Since 
$\langle \dots \rangle$ denotes the (time ordered) 
{\it grand canonical} average 
with respect to $Q$, this procedure allows the
use of the full machinery of quantum field theory, including Wick's theorem.
It is worth pointing out that the representation 
Eq.\ (\ref{eq:sbrepresentation}) with the constraint $Q=1$ enforced
by Eq.\ (\ref{eq:projection}) is exact.
Physically observable quantities are necessarily given by two
pseudoparticle (or higher) correlation functions, which in principle
requires the calculation of both, self-energy and vertex corrections.

Inspection of the terms of perturbation theory shows that each contour 
integral along the branch cut of a pseudoparticle propagator carries a 
fugacity factor $e^{-\beta\lambda}$ and, thus vanishes in the limit 
$\lambda\to\infty$. As a consequence, any bubble diagram consisting only
of pseudoparticle propagators vanishes in the physical subspace $Q=1$,
and any conduction electron propagator $G_{c\sigma}(\omega)$ appearing 
within an impurity diagram is not renormalized by the hybridization.
Only in an expectaion value of impurity operators like 
Eq.\ (\ref{eq:projection}), one factor $e^{-\beta\lambda}$ is 
cancelled by a corresponding factor in 
$\langle Q e ^{-\beta[S+\lambda(Q-1)]}\rangle$,
leaving a single pseudoparticle bubble. Details of the evaluation of 
pseudoparticle diagrams can be found in Ref.\ \cite{Costi96}.
   
\subsection{Exact properties of the pseudoparticle propagators}
The pseudoparticle propagators $G_{x}(\omega)$, $x=f\sigma$, $b$, $a$,
are given in terms of the respective selfenergies 
$\Sigma_{x}(\omega)$ as
\begin{eqnarray}
G_{x}(\omega ) &=& [G_{x}^0(\omega )^{-1} -
\Sigma _{x}(\omega)]^{-1} \label{eq:Gf}\ ,
\end{eqnarray}
where the unrenormalized propagators read, after projection,
$G_{f\sigma}^0(\omega )=[\omega - (\sigma +1)B/2] ^{-1}$,
$G_{b}^0(\omega )=[\omega +\epsilon_d - B/2 ]^{-1}$, and
$G_{a}^0(\omega )=[\omega - B/2 -U]^{-1}$.
We now derive their exact low-energy behavior.
Rewriting Eq.\ (\ref{eq:projection}) in terms of pseudoparticle 
operators and using Wick's theorem, it is seen that in $G_{x}(\omega)$
the average $\langle \dots\rangle$ is confined to the $Q=0$ sector
because of the limit $\lambda\to\infty$. This implies that the
pseudoparticle propagators have only forward in time propagation.
Therefore, these propagators are similar to the propagators 
of the X-ray threshold problem, i.e. their spectral functions
$A _{x} (\omega) = - \frac{1}{\pi} {\rm Im} G _{x} (\omega +i0) $
show infrared threshold powerlaw behavior,
\begin{eqnarray}
\label{eq:spectralfunction}
A _{x} (\omega) \sim \Theta (\omega ) \omega ^{-\alpha _x}\ , \quad
x=f\sigma,\ b,\ a \quad \omega \ll T_K \ .
\end{eqnarray}
The exact infrared exponents $\alpha _x$ can be determined for the
spin-screened FL case by the
observation that in the X-ray problem they are related to the 
scattering phase shift $\delta _\sigma^{(x)}$ via 
$\alpha _x = 1- \sum _{\sigma}\left( \delta_\sigma^{(x)}/\pi\right)^2$
and by employing the Friedel sum rule \cite{Menge88,Kroha97}. 
It links the conduction electron
number in channel $\sigma$, attracted from 
infinity in order to screen the impurity, to the phase shift,
$\Delta n_{c\sigma}^{(x)} = \delta _{\sigma}^{(x)}$.
$\Delta n_{c\sigma}^{(x)}$ in turn is defined as the difference between
the average impurity occupation number $n_{d\sigma}$ in the stationary
limit (time $t\to\infty$) and the impurity charge created by the
respective pseudoparticle operator, $f_\sigma^{\dagger}$,
$b^{\dagger}$, or $a^{\dagger}$ at $t=0$. Hence, we have for
$G_{f\sigma}$: $\Delta n_{c\sigma}^{(f)}= n_{d\sigma}-1$,
$\Delta n_{c\sigma '}^{(f)}= n_{d\sigma '}$, $\sigma ' \neq \sigma$;  
for $G_{b}$: $\Delta n_{c\sigma}^{(b)}= n_{d\sigma}$,  
and for $G_{a}$: $\Delta n_{c\sigma}^{(a)}= n_{d\sigma}-1$, $\sigma =\pm 1$.  
This implies the exact values for the infrared exponents \cite{Costi94},
valid in the low-frequency regime $\omega < T_K$,
\begin{eqnarray}
\label{eq:exponentsf}
\alpha_{f\sigma}&=&
n_d- n_d^2/2 +\Delta n_{d\sigma}(B) -\Delta n_{d\sigma}(B)^2/2 \\
\label{eq:exponentsb}
\alpha_{b}&=&1- n_d^2 /2 - \Delta n_{d\sigma}(B)^2 /2  \\
\label{eq:exponentsa}
\alpha_{a}&=&-1+2 n_d- n_d^2/2 - \Delta n_{d\sigma}(B)^2/2 \ , 
\end{eqnarray}
where $n_d =\sum _{\sigma} n_{d\sigma}$ is the total impurity occupation
number and $\Delta n_{d\sigma} (B) = n_{d\sigma} - n_{d-\sigma}$ the difference
between the $\sigma = 1$ and $\sigma = -1$ occupation numbers in a 
magnetic field. The dependence of the exponents on 
$n_d$ and $\Delta n_d (B)$  is
characteristic for FL behavior because of the use of the
Friedel sum rule (i.e. potential scattering only)
in their derivation. Therefore, whether or not the
exponents are reproduced by a given approximation can serve as an
indicator of its correctness, telling whether the approximation 
captures the spin-screened FL fixed point of the 
single-channel Anderson model. 
In the following we will focus mostly on the case of an infinite Coulomb
repulsion $U$, where the more than singly occupied states 
($a$, $a^{\dagger}$) do not contribute.

\begin{figure}[t]
\begin{centering}
\includegraphics[width=  0.95 \linewidth]{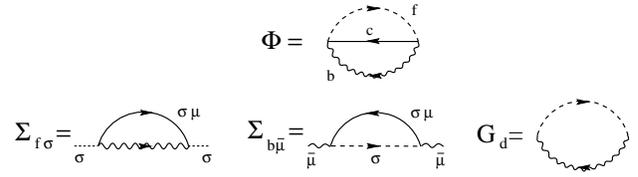}
\caption{\label{Phi_NCA} 
Diagrammatic representation of the
Luttinger-Ward functional generating the NCA for $U\to\infty$  
and the corresponding
pseudoparticle selfenergies $\Sigma _{f\sigma}$,
$\Sigma _{b}$ and the impurity electron Green's function
$G_{d\sigma}$. 
Solid, dashed, and wiggly lines represent the conduction electron,
the pseudofermion and the slave boson propagators, respectively.
Throughout this paper, all lines are understood as the renormalized
propagators, unless stated otherwise. 
}
\end{centering}
\end{figure}
\subsection{Conserving approximations}
The solution to problem (b) requires the selection and summation of
the essential terms of perturbation theory, in pseudoparticle
representation. Precondition for the projection onto the physical 
subspace is the conservation of the local charge $Q$, implied by the
symmetry with respect to simultaneous U(1) gauge transformations of the
pseudoparticles. Any approximation conserving
the projection has to preserve gauge symmetry. This is achieved by
constructing conserving approximations from a generating
Luttinger-Ward functional $\Phi$ \cite{Baym6162}. The pseudoparticle
selfenergies $\Sigma _x$ are obtained as functional derivatives of
$\Phi$ with respect to the corresponding Green's functions $G_x$, and
are thus functionals of the dressed $G_x$. In this way a closed set of
nonlinear coupled integral equations for the $G_x's$ is obtained for any
choice of $\Phi$. The choice of $\Phi$ furthermore dictates the 
calculation of the physical impurity electron Green's function 
$G_{d\sigma}$ (and other correlation functions), 
since, by definition, $|V|^2G_{d\sigma} =
\lim _{\lambda\to\infty} \delta \Phi /
\delta G_{c\sigma}$ is the single-particle conduction electon 
T-matrix.

\section{Non-Crossing Approximation}
\label{3}

In view of the small parameter $V/D$ it appears reasonable to start
with the lowest (2nd order in $V$) approximation for $\Phi$. Since in
this approximation, which sums up infinitely many self-energy
insertions, the corresponding Feynman diagrams do not have any
crossings, it has been termed the ``Non-Crossing Approximation''
(NCA) (see Fig.\ \ref{Phi_NCA}).
The NCA has been pioneered by Keiter and Kimball using the
resolvent operator formalism \cite{Keiter71,Grewe81} and by Kuramoto,
who first recognized the conserving nature of the 
NCA \cite{Kuramoto83}.
It took more than 10 years, before the NCA equations were numerically
evaluated \cite{Kojima84,Bickers87} and furthermore solved
analytically in the limit of low energies at $T=0$
\cite{Mueller8485}.  Most of these works were concerned with the limit
of infinitely strong Coulomb interaction $U\rightarrow \infty$, in
which double occupancy of the impurity is strictly excluded.  
In this case, the selfconsistent NCA equations in conjunction 
with Eq.\ (\ref{eq:Gf}) read (compare Fig. \ref{Phi_NCA}),
\begin{eqnarray}
\Sigma_{f\sigma}^{NCA}(\omega)\hspace*{-0.26cm}&=&\hspace*{-0.26cm}\Gamma\int 
              \frac{{d}\varepsilon}{\pi}\,
               f(\varepsilon )
              A_{c\sigma}^0(-\varepsilon)G_{b}(\omega +\varepsilon )
              \label{sigfNCA}\\
\Sigma_{b}^{NCA}(\omega )\hspace*{-0.26cm}&=&\hspace*{-0.26cm}
              \Gamma\sum _{\sigma}\int 
              \frac{{d}\varepsilon}{\pi}\,
              f(\varepsilon )A_{c\sigma}^0(\varepsilon)
              G_{f\sigma}(\omega +\varepsilon )
              \label{sigbNCA}\\
G_{d\sigma}^{NCA}(\omega )
         \hspace*{-0.26cm}&=&\hspace*{-0.26cm} 
         \int  {d}\varepsilon\,  {\rm e}^{-\beta\varepsilon}
         [ G_{f\sigma}(\omega +\varepsilon )A_{b}(\varepsilon )
          \nonumber\\
         &\ &\hspace*{1.5cm}-A_{f\sigma}(\varepsilon )
                   G_{b}(\varepsilon -\omega ) ] \ ,
          \label{gdNCA}
\end{eqnarray}
where $\omega$ is understood as $\omega\pm i0$ for the
retarded/advanced functions,
$\Gamma = \pi N_0 |V|^2$, and $A_{c\sigma}^0=\frac{1}{\pi}\, 
{\rm Im}G_{c\sigma}^0/N_0$ is the local
conduction electron density of states per spin,
normalized to the density of states at the Fermi level $N_0$, 
and $f(\varepsilon )=1/({\rm exp}(\beta\varepsilon )+1)$ is the 
Fermi distribution function. For $U\to\infty$
the NCA captures correctly the Kondo energy scale, and it provides a
qualitative description of the formation of the Kondo resonance.
However, it fails in a magnetic field even in the high temperature
regime $T\gg T_K$, producing a spurious resonance in the impurity
spectrum at $\omega =0$,
in addition to the two Zeeman-split Kondo peaks.
The origin of this failure will be traced in the appendix by means 
of an RG analysis.
The NCA fails furthermore at temperatures $T \ll T_K$, 
where spurious infrared singular
behavior in physical quantities appears, 
in contradiction to the expected FL behavior
\cite{Bickers87,Mueller8485,Costi96}.  
The infrared exponents of the auxiliary particle propagators come out
independent of $n_d$, $\alpha _f^{NCA} = 1/(N+1)$, 
$\alpha _b^{NCA} = N/(N+1)$, with $N$ the spin degeneracy,
again in contrast to the FL behavior
Eqs.\ (\ref{eq:exponentsf}--\ref{eq:exponentsa}).  
The low-$T$ failure of NCA is less pronounced
in SU($N$) symmetric models with $N \gg 1$. NCA becomes formally exact
in the limit $N \rightarrow \infty$, with deviations appearing in
$O(\frac{1}{N^2})$ \cite{Bickers87,Cox93}.
Note that the deviation of the approximate NCA values for the
pseudoparticle exponents $\alpha_f,b$   is of order  $1/N$,  
not  $1/N^2$  as one may have expected.

At finite Coulomb
interaction $U$, the exchange interaction $J$ acquires contributions
from both, virtual excitations to the empty and to the doubly occupied
impurity states.  A simple generalization of NCA to this case,
i.e. adding the second order perturbation theory for the two
processes, fails to capture the simultaneous contribution of both
channels in each order of bare perturbation theory, and therefore
leads to a possibly by orders of magnitude incorrect value of $T_K$
(in the worst case it is off by a factor of $T_K/D$).  An infinite
summation of vertex corrections (the symmetrized finite U NCA, or
SUNCA) is necessary to cure this problem \cite{Haule01}.

The failure of the NCA at temperatures $T \ll T_K$ can be traced back
to its insufficient inclusion of coherent multiple spin-flip
processes. The latter are responsible for the formation of the Kondo
resonance state.  A qualitative improvement therefore requires to 
include the proper vertex corrections which account for the dominant
spin and charge fluctuations. There is strong evidence that this is 
achieved by
the so-called Conserving T-Matrix Approximation (CTMA) 
\cite{Kroha97,Kroha03}. 
Before we discuss the CTMA in section \ref{6}, we will in the
next two sections present relatively easily tractable
generalizations of the NCA to adapt problems with several local
orbitals and with a large, but finite Coulomb repulsion $U$.

\section{Multi-orbital Anderson impurity systems}
\label{4}
In contrast to analytical and numerical exact solution methods, the
conserving technique is generalizable in a straight-forward way for
impurity models with several local orbitals. Such problems arise, for
instance, in
cluster extensions of the DMFT \cite{Maier00,Kotliar01}, but also in 
transition and rare earth metal impurity systems. To give an 
understanding of the wealth of the low-energy spectra of such systems, we
here treat, on the level of NCA, a Ce impurity embedded in a metallic host,
for which experimental photoemission spectra are available 
\cite{Reinert01}.
Although the compound CeCu$_2$Si$_2$ is a heavy fermion lattice system, 
for which one expects in the lattice-coherent 
low-$T$ state a dispersion of the Kondo quasiparticle resonance with
width of order of $T_K$, the photoemission measures the momentum integrated
spectral density (implying only 
a slight broadening of the observed resonance), so that direct 
comparison with the results of the impurity model are possible.

Ce has seven 4f orbitals, which are spin-orbit (SO) and crystal-field (CF)
split and have an overall valence of close to 1 in CeCu$_2$Si$_2$.
Hence, one can assume a large on-site Coulomb repulsion $U\to\infty$
between all 4f orbitals, and the model Hamiltonian reads in auxiliary
particle representation,
\begin{eqnarray}
\label{eq:hamiltonianCe}
H&=& \sum_{\vec{k} \sigma} \epsilon_{\vec{k}} 
c^{\dagger}_{\vec{k} \sigma}c^{ }_{\vec{k} \sigma}\,+\, 
\sum_{m\sigma} \epsilon_{dm}
f_{m\sigma}^{\dagger}f_{m\sigma}^{ }  \\
&+&  \sum_{ m 
\vec k \sigma }V_m (c^{\dagger}_{\vec k \sigma } b^{\dagger}f_{m\sigma}^{ } 
+ h.c.) \ , \nonumber 
\end{eqnarray}
where  $\epsilon_{dm}$ are the energies of the 4f orbitals
and $V_m$ the hybridization matrix element of the local orbital $m$
with the conduction band, $m=1,\dots, 7$. 
The resulting generalized NCA equations follow as,
\begin{eqnarray}
\hspace*{-0.4cm}&\Sigma&\hspace*{-0.4cm}_{fmm '\sigma}(\omega)=
             \Gamma _{mm'}\int 
              \frac{{d}\varepsilon}{\pi}\,
               f(\varepsilon )
              A_{c\sigma}^0(-\varepsilon)G_{b}(\omega +\varepsilon )
              \phantom{xxxxii}
              \label{sigfNCAm}\\
&\Sigma&\hspace*{-0.4cm}_{b}(\omega )
              =
              \sum _{mm '\sigma}\hspace*{-0.1cm}\Gamma_{m 'm }\hspace*{-0.1cm}
              \int 
              \frac{{d}\varepsilon}{\pi}\,
              f(\varepsilon )A_{c\sigma}^0(\varepsilon)
              G_{fmm '\sigma}(\omega +\varepsilon )
              \label{sigbNCAm}\\
&G&\hspace*{-0.4cm}_{d mm '\sigma}(\omega )
         =
         \int  {d}\varepsilon\,  {\rm e}^{-\beta\varepsilon}
         [ G_{fmm '\sigma}(\omega +\varepsilon )A_{b}(\varepsilon )
          \nonumber\\
         &\ &\hspace*{3cm}-A_{fmm '\sigma}(\varepsilon )
                   G_{b}(\varepsilon -\omega ) ] \ ,
          \label{gdNCAm}
\end{eqnarray}
\begin{figure}[t]
\begin{centering}
\includegraphics[width=  0.8 \linewidth]{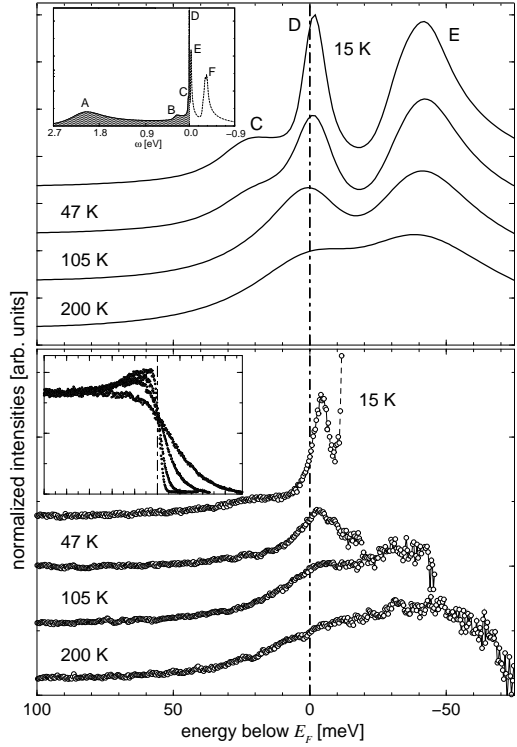}
\caption{\label{NCA_PES}
Upper panel: 
      Theoretical $T$ dependence of the 4f
      spectral function of CeCu$_2$Si$_2$ for 
      $T=15$~K, $47$~K, $105$~K, and $200$~K.
      The inset shows the calculated spectrum at $T=15\ K$.
      Model parameters: $\epsilon_{f1}=-2.4$~eV, CF
      splittings of the $J=5/2$ sextet $\Delta_{CF}=30$~meV and $36$~meV, SO
      splitting $\Delta_{SO}=270$~meV, hybridization $V=200$~meV.
Lower panel: Photoemission spectra for the same temperatures.
      The experimental photoemission spectra, 
      divided by the Fermi distribution to
      gain access to the states above the Fermi energy.
      The inset shows the raw data on the same energy scale
      prior to the division. All spectra are normalized to the same
      intensity at $\approx100$~meV and are offset for clarity. 
}
\end{centering}
\end{figure}
where $\Gamma_{mm '}=\pi N_0 V_m^*V_{m '}$, and $G_{fmm '\sigma}$,
$G_{dmm '\sigma}$
are the matrix generalizations of Eq.\ (\ref{eq:Gf}) 
associated with the selfenergy matrix $\Sigma_{fmm '\sigma}$ in 
local orbital space.
The fact that there are several single-particle levels at high
energies, $\epsilon _{fm}$, grouped together in SO and CF multiplets,
implies a rich structure in the low-energy spectrum as well, because
fluctuations of an electron from one local orbital via the conduction band
to another local orbital are possible in second order in $V_m$. Since
there may be a spin flip involved, these fluctuations induce Kondo-like,
logarithmic divergencies in perturbation theory, however located at
energies corresponding to the {\it differences} $\Delta \epsilon _{m1}= 
\epsilon _{fm}-\epsilon _{f1} >0$ between an excited and the ground state
local level, as can be seen, e.g., by inserting the bare pseudofermion
Green's function $G_{fm m \sigma}$ in Eq.\ (\ref{sigbNCAm}). As
a result, there appear (in addition to the central Kondo resonance at the
Fermi level), multiple many-body resonances at elevated 
energies $\Delta \epsilon _{m1}$ given by the SO and CF splittings,
as well as shadow peaks at negative energies
$-\Delta \epsilon _{m1}$, the so-called SO and CF satellites.
The height of these resonances depends roughly logarithmically on $T$, 
characteristic of Kondo behavior. In addition, the SO and CF
satellites are broadened by the lifetime associated with the
inelastic decay of the excited local orbitals. The shadow peaks 
often appear merely as shoulders in the spectrum, as they correspond
to transitions from an only virtually occupied excited 4f orbital
into the ground state orbital. 
This physics has been analyzed analytically in Ref.\ \cite{Kroha02}
and is qualitatively well captured by the NCA, as a perturbative expansion of
Eqs.\ (\ref{sigfNCAm}--\ref{gdNCAm}) shows.
The results of the numerical evaluation of the NCA equations 
for the impurity spectrum $\frac{1}{\pi}{\rm tr}_m \left[ {\rm Im} 
G_{d\sigma} (\omega -i0)\right] $ are shown in
Fig.\ \ref{NCA_PES} in comparison to the photoemission spectra of
Ref.\ \cite{Reinert01}. The central Kondo Peak D and the CF satellites 
(peaks E and shoulders C)
with  CF splittings of 30 meV and 36 meV are clearly visible in the main 
panels. The inset of the upper panel displays the NCA spectrum on a larger
energy scale, showing the SO satellites (B and F) with a SO splitting $\approx$
360 meV as well as the single-particle resonances A, whose SO and CF   
splittings are not resolved due to the large lifetime broadenings 
$\Gamma _{mm '}$. The $T$ dependence agrees qualitatively well with
the photoemission spectra in the experimentally accessible temperature
range of $T\stackrel{>}{\sim} T_K$.

\begin{figure}[b]
\begin{centering}
\includegraphics[width=  0.95 \linewidth]{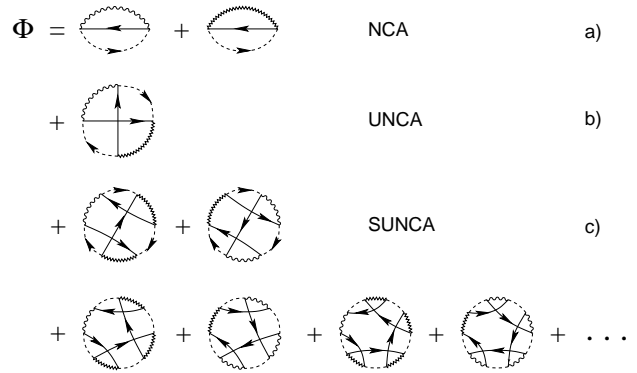}
\caption{\label{PhiSUNCA} 
Generating functional of the SUNCA. Zig-zag lines
denote the propagators for doubly occupied or ``heavy'' bosons.
The diagrams a) define the NCA for finite $U$,
the sum of diagrams a) and b) the UNCA, which gives a symmetrical
treatment of b and a lines to leading order only.     
}
\end{centering}
\end{figure}
\section{Finite-$U$ Anderson impurities: SUNCA}
\label{5}
\begin{figure}[t]
\begin{centering}
\includegraphics[angle=-90, width=0.9\linewidth]
{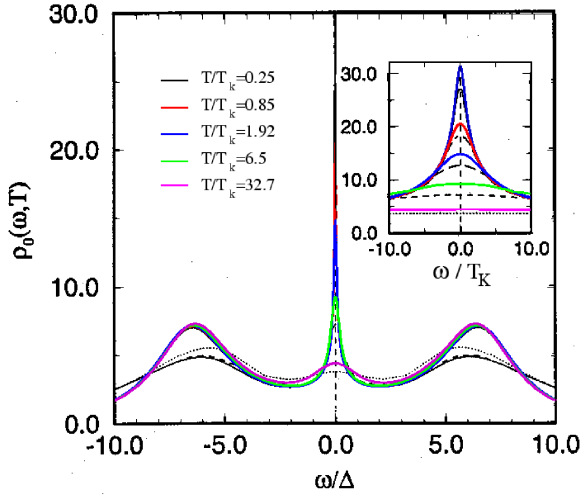}
\caption{\label{Ad_NRG_SUNCA} 
$d$-electron spectral function for 
$U=-2\epsilon _d$. Solid lines: SUNCA results \cite{Haule01}, 
dashed lines: NRG results \cite{Costi}.
}
\end{centering}
\end{figure}

For many quantum impurity problems, including the DMFT treatment of
correlated lattice electron systems with a Mott-Hubbard metal-insulator
transition, the large, but finite on-site Coulomb repulsion $U$ is essential.
However, a naive generalization of the NCA for finite $U$, the generating
functional comprised of the two diagrams in the first line of 
Fig.\ \ref{PhiSUNCA}, does not even give the correct Kondo scale for
this problem. The reason is that in this NCA the fluctuations into the
empty (``light'' bosons $b$) and into the doubly occupied 
(``heavy'' bosons $a$) impurity state are not treated symmetrically.
The latter is, however, essential to obtain the correct spin exchange
coupling
\begin{eqnarray}
\label{JfiniteU}
J=\frac{|V|^2}{|\epsilon _d|}+\frac{|V|^2}{|\epsilon _d +U|} 
\end{eqnarray}
via a Schrieffer-Wolff transformation \cite{Schrieffer66}.
Thus, the asymmetric treatment of either of the two contributions 
to $J$ leads to a $T_K$ which may be off by an exponentially large
factor $\sim e^{\pi |\epsilon _d|/2\Gamma}$ or
$\sim e^{\pi |\epsilon _d +U|/2\Gamma}$, i.e. possibly by 
several orders of magnitude.
For a correct treatment of both terms, there must be included,
for each diagram with a light boson line, the corresponding diagram
with the light boson replaced by a heavy boson line (which amounts to
the exchange diagram of the former), and vice versa,
{\it on the level of bare perturbation theory} \cite{Haule01,Holm89}. 
The importance of these vertex corrections has first been recognized 
by Sakai et al. \cite{Sakai88} and later by Pruschke and 
Grewe \cite{Pruschke89}, without, however, formulating a
conserving approximation. In these works a numerical evaluation has
only been given in lowest order (the terms denoted UNCA in Fig.\ \ref{PhiSUNCA}).
On the level of renormalized perturbation theory, it means that for
each dressed $b$-line there must be included a 
ladder vertex function with $a$-lines as rungs, and vice versa.
The generating functional of the corresponding conserving approximation,
the so-called Symmetrized Finite-U NCA (SUNCA), is shown in 
Fig.\ \ref{PhiSUNCA} \cite{Haule01}. The SUNCA is tractable with
relatively moderate numerical effort, since it can be formulated
in terms of no higher than 3-point vertex functions.
The results of a fully selfconsistent evaluation of the
$d$-electron spectral function within 
SUNCA are shown in Fig.\ \ref{Ad_NRG_SUNCA} in comparison
with NRG results. It is seen that the correct Kondo scale 
(width of the Kondo peak) is reproduced. However, the SUNCA
solution still develops a spurious low-$T$ singularity.
This problem can be cured only by a more sophisticated 
approximation, the CTMA.

\section{Conserving T-Matrix Approximation}
\label{6}

To overcome the failures of the NCA to describe the FL 
strong coupling fixed point of the single-channel Anderson model,
described in section \ref{3}, we have proposed early on the CTMA
\cite{Kroha97}.
Before we present the explicit results of the CTMA,
we give the line of arguments that led to the construction of the
generating functional of the CTMA.

\begin{figure}[t]
\begin{centering}
\includegraphics[width=  0.90 \linewidth]{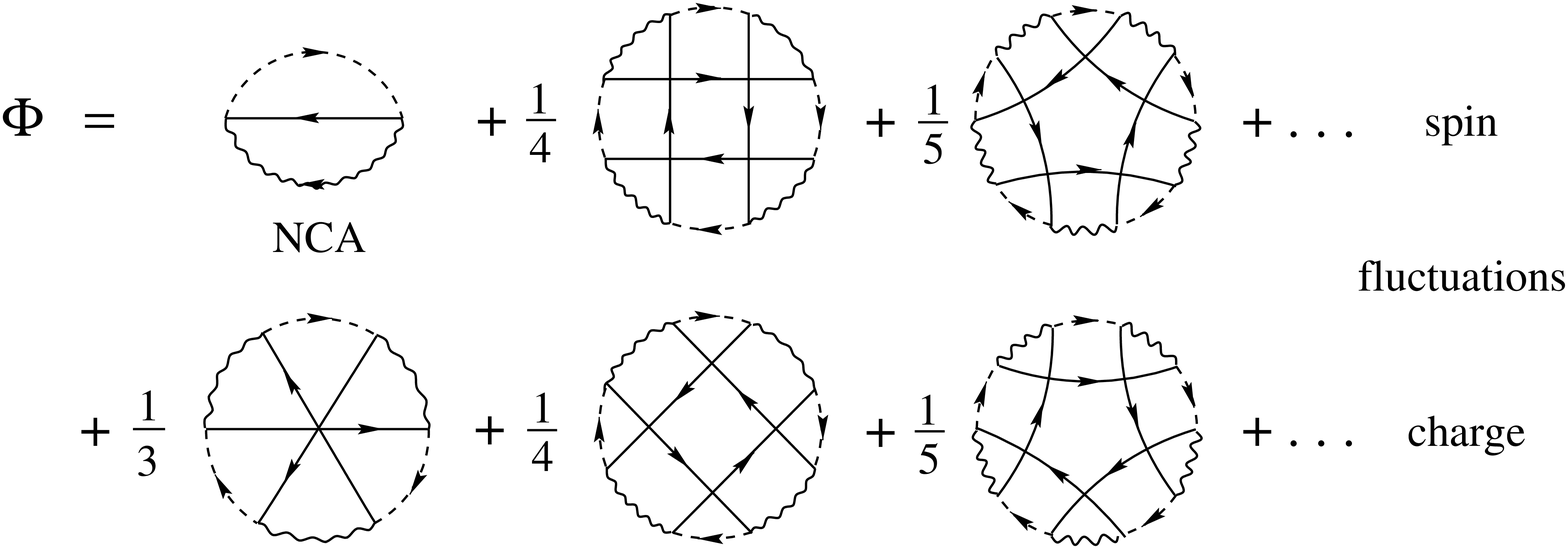}
\caption{\label{Phi_CTMA} 
Diagrammatic representation of the
Luttinger-Ward functional generating 
CTMA. The terms with the conduction 
electron lines running clockwise (labelled ``spin fluctuations'') generate 
$T^{(cf)}$, while the terms with the conduction electron 
lines running counter-clockwise (labelled ``charge fluctuations'')
generate $T^{(cb)}$ (Fig.\ \ref{vertices_4p}, see text). 
The two-loop diagram is excluded, because it is not a skeleton. 
}
\end{centering}
\end{figure}
\subsection{Construction of the CTMA}
\begin{figure}[b]
\begin{centering}
\includegraphics[width=  0.95 \linewidth]{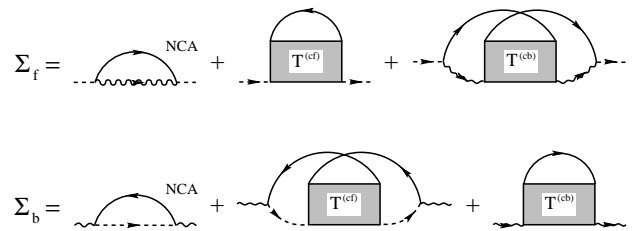}
\caption{\label{selfenergyCTMA} 
The CTMA pseudofermion and slave boson selfenergies.
In the second diagram of $\Sigma _{f\sigma}$ and the 
the third diagram of $\Sigma _{b}$ the 1- and 2-rung contributions, 
and in the third diagram of $\Sigma _{f\sigma}$ and the 
the second diagram of $\Sigma _{b}$ the 1-rung contribution 
to the respective T-matrices are omitted,  
(not shown explicitly), in order to avoid double counting.}
\end{centering}
\end{figure}

As a minimal precondition to obtain a gauge symmetric description of the
FL fixed point, a conserving approximation must reproduce the correct
pseudoparticle infrared exponents,
Eqs.\ (\ref{eq:exponentsf}-\ref{eq:exponentsa}), whose dependence on $n_d$
is characteristic for a FL ground state. It is easily seen by power counting 
arguments that any summation of a finite number of skeleton selfenergy
diagrams merely reproduces the incorrect NCA exponents \cite{Cox93}. 
Hence, the generating functional must be comprised of an 
{\it infinite} class of skeleton diagrams in order to describe FL
behavior, in contrast to the Post-NCA considered by Anders (diagrams 
up to O($\Gamma ^4$) in Fig. \ref{Phi_CTMA}) \cite{Anders94}. 
Since the latter is a consequence of the singlet state
formed at low $T$ between the impurity and the conduction
electron spins, one may expect that higher than two-particle 
correlation functions need not be considered in the single-channel
case. The approximations to the total vertex functions between
conduction electons ($c$) and impurity degrees of freedom
(pseudofermions $f$, slave bosons $b$) are then two-particle
T-matrices.  As the irreducible parts of these T-matrices
we select the single (renormalized) $b$ or $f$ particle lines,
since (1) in the Kondo regime these terms are the leading contributions 
in the small parameter $VN_0$; 
and (2), in the spirit of principal diagrams, this choice gives rise to
the maximum number of spin and charge fluctuation processes in the
T-matrices at any given order of (renormalized) perturbation theory.
This choice results in the ladder approximations $T^{cf}$, $T^{cb}$ 
for the total
two-particle vertices shown in Fig.\ \ref{vertices_4p} (2), (3).
The Luttinger-Ward functional generating these ladder vertex terms
(and others) is shown in 
Fig.\ \ref{Phi_CTMA}. It is comprised of all closed pseudoparticle rings 
(skeletons) with each conduction electron line spanning at most
two hybridization vertices and has been termed the ``Conserving T-Matrix 
Approximation'' (CTMA). 
The pseudofermion and slave boson
selfenergies, derived by functional derivative, are shown in 
Fig.\ref{selfenergyCTMA}. 
Note that the vertex equations for $T^{cf}$, $T^{cb}$, coupled via
the selfenergies, have parquet character. 
The analytical expressions for $\Sigma _{f\sigma}$, $\Sigma _b$
and $G_{d\sigma}$ are given explicitly in Ref.\ \cite{Kroha03,Kirchner04}.

The selfconsistent CTMA equations are solved numerically by 
iteration. The results shown below support strongly that the
CTMA correctly describes the strong coupling FL regime of the
single-channel Anderson impurity model. In the appendix evidence is
provided that it also systematically describes the high-energy regime
above $T_K$.

\begin{figure}[t]
\begin{centering}
\includegraphics[width=  0.8  \linewidth]{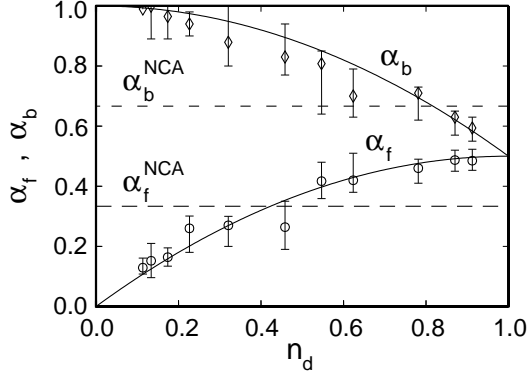}
\caption{\label{exponents} 
CTMA results (symbols with error bars) for the threshold 
exponents $\alpha _f$ and $\alpha _b$ of $A_{f\sigma}$ and $A_b$
for $U\to\infty$, $B=0$. 
Solid lines: exact values, Eqs.\ (\ref{eq:exponentsf},\ref{eq:exponentsb}), 
dashed lines: NCA results.}
\end{centering}
\end{figure}
\subsection{Results}
{\it Pseudoparticle spectral functions.} 
As a first indication for description of FL behavior within CTMA
it has been checked if the CTMA reproduces the correct FL values
of the pseudoparticle threshold exponents
\cite{Kroha97}. The exponents were obtained by a fit to the numerical 
low-$T$ solutions for $A_{f\sigma}$, $A_{b}$ in a log-log plot.
The results are shown in Fig.\ \ref{exponents}, showing, within the
error bars, good agreement with the exact values and especially the
dependence on the impurity occupation number $n_d$, characteristic
for the FL fixed point.

\begin{figure}[t]
\begin{centering}
\includegraphics[width=  0.88 \linewidth]{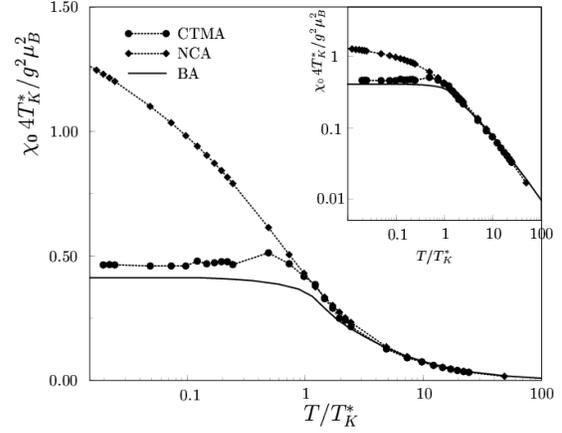}
\caption{\label{susc} 
Static spin susceptibility as a function of temperature; 
BA, CTMA and NCA results (see text). Model parameters used:
$\epsilon _d /D =-0.81$, $\Gamma /D =0.2$. 
}
\end{centering}
\end{figure}

{\it The static spin susceptibility} of the impurity
was calculated from the spin dependent occupation numbers
$n_{d\sigma}$ in a magnetic field ${\cal B}$ as,
\begin{equation}
\chi_i(T) = \left. \frac{dM}{d{\cal B}}\right|_{{\cal B}=0}
\label{eq:susc1}
\end{equation}
where $M=g\mu_B\sum_{\sigma}\sigma n_{\sigma}$ is the impurity magnetization
and
\begin{equation}
n_{d\sigma} = \lim_{\lambda \to \infty}
\frac{\int d\omega e^{-\beta\omega}\; \mbox{Im} G_{f\sigma}(\omega-i0)}
{\int d\omega e^{-\beta\omega}\; \mbox{Im}[\sum_{\sigma} 
G_{f\sigma}(\omega-i0)+ G_{b}(\omega-i0)]}\ .
\label{eq:nsigma}
\end{equation}
$\chi_i(T)$ shows $T$-independent Pauli behavior for $T\stackrel{<}{\sim}0.5
T_K$ and down to the lowest $T$ considered \cite{Kirchner04}, 
indicative of the FL ground state with a completely screened local moment
(Fig.\ \ref{susc}). 
As expected, $\chi_i(T)$ obeys scaling for at least a range of
$T_K$ within a factor $10$ \cite{Kroha_unpub}, 
when plotted as a function of $T/T_K$
and in units of $(g\mu_B)^2/(4T_K^*)\equiv \chi _i(T=0)/W$, where
$W=T_K^*/T_L$ is the universal Wilson number and 
$T_K^*$ is Wilson's original definition of $T_K$ \cite{Wilson75}. 
Since the BA
and selfconsistent perturbation theory use somewhat different definitions 
of $T_K$, we rescale the latter in the CTMA as well as in the BA solution such
that it coincides with $T_K^*$.
This allows for a quantitative comparison of BA and CTMA and for a
determination of the CTMA approximation to $W$.  
The result is shown in Fig.\ \ref{susc}, exhibiting remarkably good
quantitative agreement. We obtain $W^{(CTMA}\simeq 0.462$ as compared to the
exact result $W\simeq 0.4128$.
 
\begin{figure}[b]
\begin{centering}
\includegraphics[width=  0.95 \linewidth]{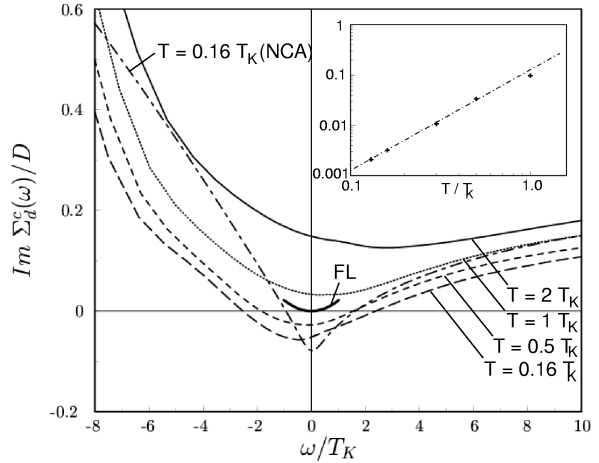}
\caption{\label{Sigmad0}
Imaginary part of the CTMA impurity electron selfenergy 
${\rm Im}\Sigma _{s\sigma}(\omega)$ for various $T$
for the same parameters as in Fig.\ \ref{susc}. The small solid arch
represents the exact FL behavior at $T=0$ \cite{Kirchner04}. 
The NCA result at $T=0.16 T_K$ is shown for comparison.
Inset: the $T$ dependence of the minimum value of 
${\rm Im}\Sigma _{s\sigma}$ (black dots). The dashed-dotted line,
drawn for comparison, has slope 2. 
}
\end{centering}
\end{figure}
\begin{figure}[t]
\begin{centering}
\includegraphics[width=  0.9 \linewidth]{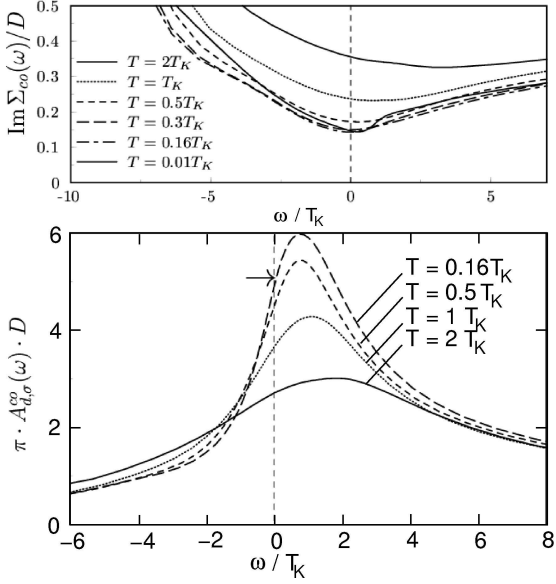}
\caption{\label{SigAd} 
$d$-electron selfenergy $\Sigma _{d\sigma}(\omega )$
and spectral function $A_d(\omega )$ after the
potential scattering correction has been applied (see text).
The arrow (lower panel) marks the unitarity sum rule.
}
\end{centering}
\end{figure}

FL behavior of the impurity electron spectral function 
$A_{d\sigma}(\omega )$ 
and selfenergy $\Sigma_{d\sigma}(\omega )$ 
is of prime interest especially for  applications
within DMFT.  
The imaginary part of the
impurity electron interaction selfenergy $\Sigma _{d\sigma}(\omega -i0)$,
calculated in CTMA \cite{Kirchner04} from 
\begin{eqnarray}
G_{d\sigma}(\omega)=[\omega -\epsilon _d -i\Gamma
-\Sigma _{d\sigma}(\omega)]^{-1} \ ,
\label{GdSigd}
\end{eqnarray}
is shown in Fig.\ \ref{Sigmad0}.
$\Sigma _{d\sigma}$ exhibits many features of FL behavior.
It has quadratic dependence on both, $\omega$ and $T$, at low $\omega$, $T$,
with no sign of a spurious low-energy singularity
down to the lowest $T$ considered ($T\simeq 0.01\ T_K$).
As discussed in detail in Ref.\ \cite{Kirchner04},
the curvature of the quadratic behavior in $\omega$ and $T$ is found to be
in good agreement with the exact FL result, 
$\Sigma _{d\sigma}(\omega) = a [\omega ^2 +(\pi T)^2]/T_K^2$,
where $a$ is an exactly known prefactor \cite{Hewson93,Kirchner04}.
However, the position $\omega _0$ of the minimum of
${\rm Im}\Sigma _{d\sigma}(\omega)$ is incorrectly shifted to 
$\omega _0 \approx - T_K$, and ${\rm Im}\Sigma _{d\sigma}(\omega -i0)$
acquires negative values, thus violating the Friedel sum rule.
When searching for the origin of this shortcoming, one must keep in
mind that $\Sigma _{d\sigma}(\omega \approx 0)$ is determined via
Eq.\ (\ref{GdSigd}) by both Im$G_{d\sigma}(\omega)$ 
and Re$G_{d\sigma}(\omega)$, and thus, through the Kramers-Kronig
relation, by high-energy (potential scattering) contributions to
$A_{d\sigma}(\omega )$. Hence, the erroneous shift $\omega _0$ may 
result from an unprecise calculation of $G_{d\sigma}(\omega)$ at
high energies, either numerically or due to neglect of high-order
potential scattering terms \cite{remark}. To correct this 
shortcoming, it has been suggested to add an appropriate,
phenomenological {\it real constant} to $\Sigma _{d\sigma}(\omega )$.
Through selfconsistency it acts like a chemical potential and 
shifts the minimum of Im$\Sigma _{d\sigma}(\omega )$ to $\omega =0$.
The results of this correction are shown in Fig. \ref{SigAd}.
It is seen that by a single, real parameter, motivated by potential
scattering contributions, the full FL behavior of $\Sigma _{d\sigma}(\omega )$
is recovered, and $A_d(\omega )$ obeys the unitarity sum rule with 
good precision.

\section{Conclusions}
We have reviewed the method of conserving auxiliary particle 
approximations for quantum impurities with strong onsite repulsion
$U$. We have shown that on the level of generalized NCA relatively
easily tractable approximations are possible which give a qualitatively
correct description 
of, e.g., Anderson impurities with multiple orbitals
or finite $U$ (SUNCA) at not too low temperature, but invariably fail
in the strong coupling regime below $T_K$ as well as in a magnetic field. 
It was shown, both by numerical solutions and by
a perturbative RG analysis, that the CTMA, although numerically 
demanding, provides a remedy for all of these
failures, and gives an essentially correct description of the FL 
behavior in the Anderson impurity model. 
It remains to be investigated if this makes the CTMA a suitable 
``impurity solver'' within the DMFT approach to correlated lattice 
problems.
 
We are grateful for useful discussions
with S. Kirchner, K. Haule, T. A. Costi, A. Rosch, H. Keiter, G. Kotliar and
G. Sellier. 
Parts of this work have been done in collaboration with
K. Haule, T. A. Costi, and, in particular, S. Kirchner.
We acknowledge the hospitality of the Aspen Center for Physics,
where this paper has been completed.
This work is supported in part by the DFG and by a Max-Planck Research
award (P.W.).

\begin{figure}[b]
\begin{centering}
\includegraphics[width=  0.95 \linewidth]{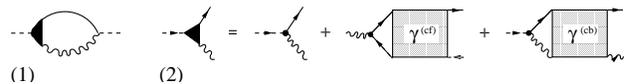}
\caption{\label{vertex_3p} 
(1) Pseudofermion selfenergy expressed in terms of the exact
3--point vertex $\hat V$ (black triangle).
(2) Representation of $\tilde V$ in terms of the 4--point vertices 
 $\gamma ^{cf}$, $\gamma ^{cb}$. The first diagram on the 
right--hand side is the bare 3--point vertex $V$. The 
pseudofermion-slave boson vertex $\gamma ^{fb}$ vanishes 
by the projection onto $Q=1$.
}
\end{centering}
\end{figure}
\appendix
\section{Perturbative RG analysis}

As seen in section \ref{6}, conserving approximations
provide a means for selecting the physically essential contributions even in
the strong coupling regime, while preserving the symmetries of the
problem. However, because of the selfconsistent resummation it is 
non-trivial to find an approximation which also includes the leading
logarithmic terms at each order of {\it bare} perturbation theory.
The latter is known to provide a quantitatively
correct description of physical quantities in the
weak and intermediate coupling regime, i.e. as long as
${\rm ln} ({\rm max}[T,\omega]/T_K) \stackrel{>}{\sim}  1$. 
On the other hand, the leading log summation 
is equivalent to the perturbative coupling constant RG in the sense
specified below. For finding the proper conserving approximation 
it is, therefore, useful to have a method which allows to analyze 
whether or not the approximation reproduces the
correct weak coupling RG flow of the effective coupling constants
under high-energy cut-off rescaling. 
In this appendix we present
such a method and apply it to the NCA and to the CTMA for the
Anderson model in the Kondo regime for $U\to\infty$.

The coupling constant RG starts from the observation that in a 
renormalizable system, like
the Kondo problem, where the $T$, $\omega$ and $B$ dependence of
physical quantities $\chi$ is characterized by a single low-energy scale
$T_K$ (i.e. given by a universal function of the dimensionless variables
$T/T_K$, $\omega/T_K$, $B/T_K$ etc.), these quantities must be invariant
under high-energy cut-off rescaling $D\to D-dD$, and so must the
two-particle vertex functions, the physical quantities are comprised
of. Hence, the direct calculation of $\chi$ is replaced by the 
renormalition of the vertex operators $\Lambda$ of the Hamiltonian such as to 
keep the total (fully reducible) vertex functions 
$\gamma ^{cf}$ (spin and potential scattering) and 
$\gamma ^{cb}$ (potential scattering) invariant under cut-off rescaling,
and subsequent calculation of $\chi$ to leading order bare perturbation theory
in the renormalized couplings. 
The same RG scheme can be applied, if $\chi$ is not given exactly, but
in some conserving approximation. In this case, one must identify the
approximate $\gamma ^{cf}$, $\gamma ^{cb}$ as dictated by the approximation
for $\chi$. Hence, the RG analysis of a conserving approximation amounts
to the following procedure \cite{Kirchner02}, see also \cite{Schiller01}:

(1) {\it Determination of the approximate vertex functions.} In any
approximation, a physical quantity can be expressed in terms of the
full hybridization vertex $\tilde V$ and of {\it bare} Green's functions,
the latter being RG invariant by definition, e.g. the $f$
selfenergy (Fig.\ \ref{vertex_3p} (1)),
\begin{eqnarray}
\Sigma _{f\sigma} (\omega ) &=& 
V \int \frac{d\varepsilon}{2\pi} f(\varepsilon) G^0_b(\varepsilon +\omega -i0)
\times
\label{sigmafexact}\\
&\phantom{+}&\hspace*{-2.3cm}
\big[ \tilde V (\omega ,\varepsilon +i0)\, G^0_{c\sigma} (-\varepsilon -i0)-
   \tilde V (\omega  ,\varepsilon -i0)\, G^0_{c\sigma} (-\varepsilon +i0) 
\big] .
\nonumber
\end{eqnarray}
$\tilde V$, in turn, is composed of four-point vertex functions
as shown in Fig.\ \ref{vertex_3p} (2). This defines the approximate 
$\gamma ^{cf}$, $\gamma ^{cb}$.
Note that only for a conserving approximation this definition 
is unique, independent of the physical quantity considered. \newline 
(2) {\it Bare perturbation theory for $\gamma ^{cf}$, $\gamma ^{cb}$.}
Due to the conserving method, the four-point vertex functions are comprised
of the dressed propagators. In order to perform the perturbative RG,
these must be written out in bare perturbation theory (non-skeleton
diagrams). In leading log approximation, 
there are two significant simplifications: (i) Each contribution to the
pseudofermion selfenergy contains at least one integration over the 
slave boson propagator $G_b$ (compare, e.g. Eq.\ (\ref{sigfNCA})).
Its logarithmic contribution is cut off at high energies $\omega \gg T_K$ 
by the pole in $G_b(\omega)$
at $\omega \approx \epsilon _d$. Note that this remains true
even for the dressed $G_b$.
Hence, 
only the bare pseudofermion Green's function appears in 
$\gamma ^{cf}$, $\gamma ^{cb}$. 
(ii) Selfenergy insertions to propagators which are integrated over
in $\gamma ^{cf}$, $\gamma ^{cb}$ are of subleading order and should,
therefore, be neglected. \newline
(3) Perform on the terms obtained in this way the perturbative cut-off
rescaling scheme \cite{Anderson70,Hewson93}
to obtain the RG equations for the spin and potential
scattering coupling constants. 

\subsection{Perturbative RG for the NCA}
\begin{figure}[t]
\begin{centering}
\includegraphics[width=  0.92\linewidth]{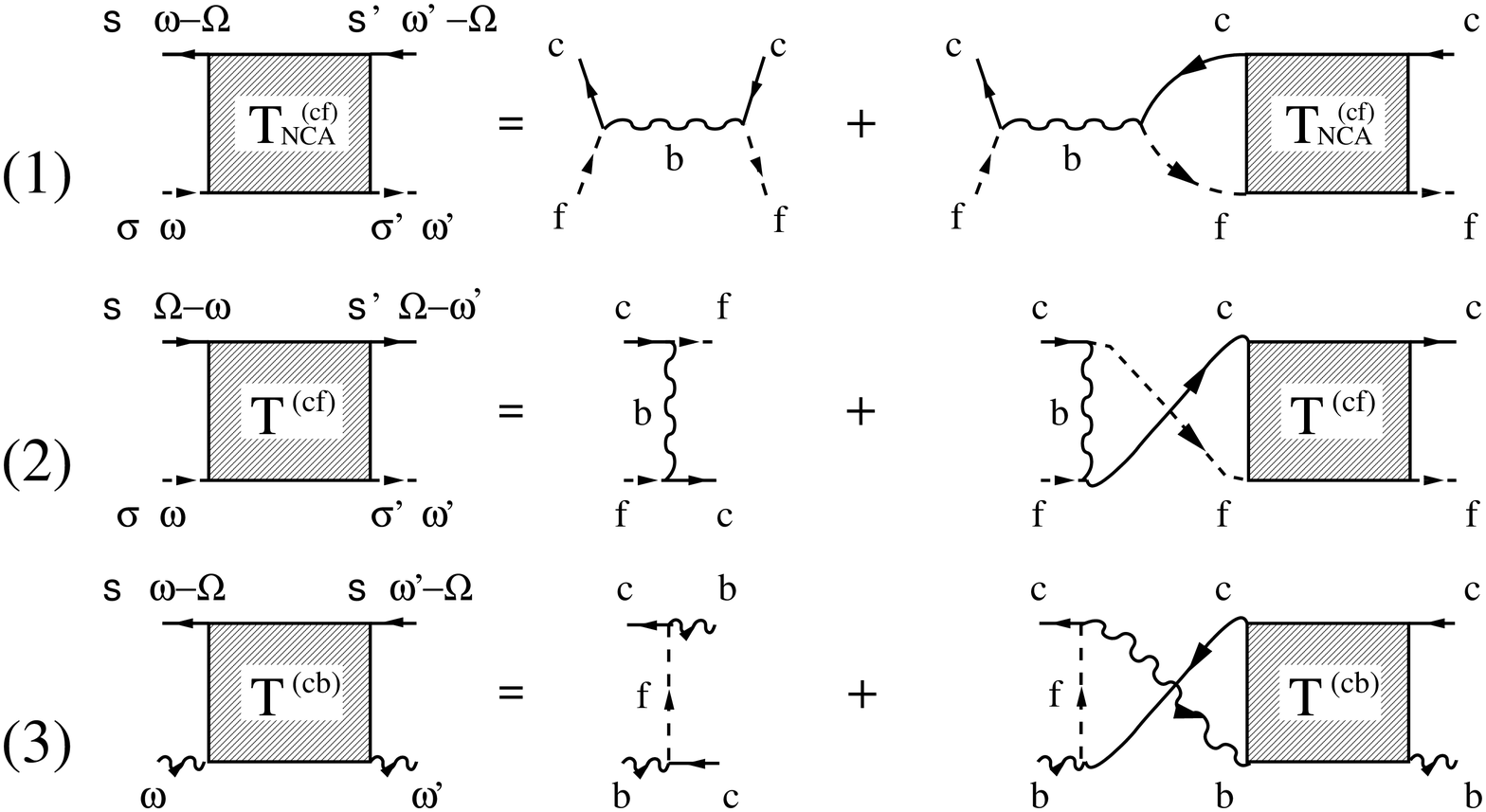}
\caption{\label{vertices_4p} 
Diagrammatic representation of the Bethe--Salpeter equations for (1) 
the NCA approximation to the 
fullly reducible conduction electron--pseudofermion p--h vertex 
$\gamma^{(cf)}_{NCA}$ 
(2) the $c$--$f$ 
T-matrix $T^{cf}$, and (3) the $c$--$b$ 
T--matrix $T^{cb}$. The CTMA approximation to 
$\gamma^{cf}$ is the sum of (1) and (2). 
For the perturbative RG, all single-particle
propagators are understood to be the bare ones (see text).
The external lines 
are drawn for clarity and do not belong to the vertices.
Single-particle
reducible parts (SPRs) in (2) are omitted, since
the SPRs of the total $c$-$f$ vertex do not contribute to the RG flow because
of rule (i) (see text).
}
\end{centering}
\end{figure}

\begin{figure}[t]
\begin{centering}
\includegraphics[width=  0.95 \linewidth]{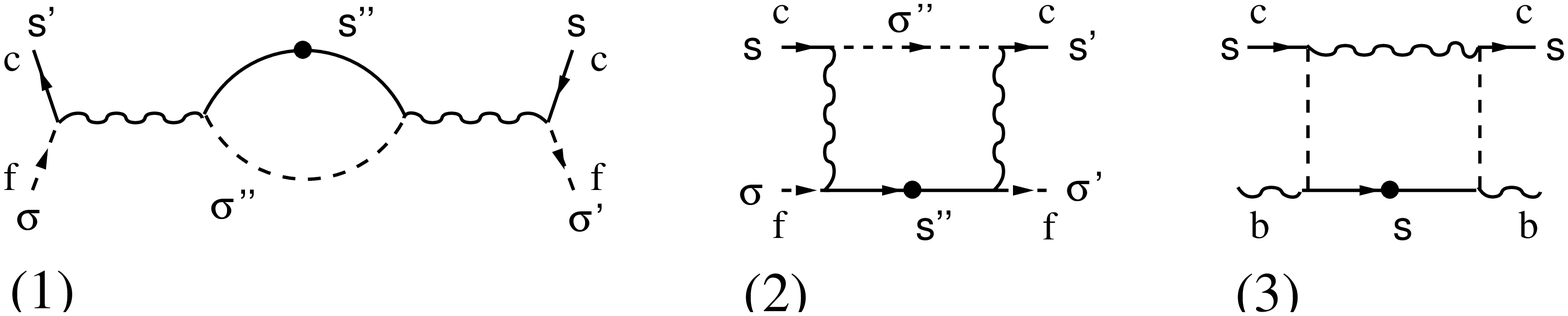}
\caption{\label{rgcorrections} 
Perturbative RG renormalizations of the (irreducible) c--f vertex.
Diagram (1) is the NCA result, the sum of diagrams (1), (2), (3) is the
result of CTMA, where (3) is the contribution from the c--b vertex,
which is not scaling and is neglected. 
The black dots on the conduction electron lines
indicate that the frequency integrals in those lines are 
restricted to the regions $[-D,-D+dD]$, $[D,D-dD]$.  
}
\end{centering}
\end{figure}

Considering the NCA selfenergies in Fig.\ \ref{Phi_NCA}
(or the NCA d-electron Green's function) and following the steps (1),
(2) above, $\gamma ^{cf}$ is identified as the full
slave boson propagator, $\gamma ^{cf}_{s' \sigma ',s \sigma}
=|V|^2 G_b$, i.e. by the
expression given diagrammatically by Fig.\ \ref{vertices_4p} (1). 
The conduction electron-slave boson vertex function $\gamma ^{cb}$
is by construction proportional to the unit matrix in spin space
since the slave bosons don't carry spin,
and can only give potential scattering contributions. 
Within NCA, it is $\gamma ^{cb}_{s' ,s } = |V|^2 G_{f\sigma} 
\delta _{ss '}\delta _{ss '}$, i.e. of subleading order. 
Performing the cut-off rescaling $D \to D-d D$ 
on $\gamma ^{cf}$ , one arrives,
to first order in $d D$ and in leading logarithmic approximation, 
at the correction to the irreducible c-f vertex, 
\begin{eqnarray}
d\Lambda ^{cf, (1)}_{s' \sigma ',s \sigma}
                 = - N_0 \frac{dD}{D} \sum _{s'' \sigma ''}
                    \big[ \Lambda^{(cf)} \big]_{s' \sigma '',s''\sigma}
                    \big[ \Lambda^{(cf)} \big]_{s''\sigma ' ,s  \sigma''}\ ,
\nonumber
\end{eqnarray}
with the spin indices as defined in Fig.\ \ref{rgcorrections} (1). 
The dot in Fig.\ \ref{rgcorrections} indicates
that the integration over the conduction electron energy is restricted to
the infinitesimal range $[-D,-D+d D]$, $[D-d D,D]$. 

$\Lambda ^{cf}$, 
$\delta\Lambda ^{cf}$ are  matrices in the product space of 
the pseudofermion and conduction electron spins, and, hence, can be
decomposed into spin flip, 
$J_{\perp}\ (S^+\otimes\sigma^- + S^-\otimes\sigma^+)$,
spin z-component, $J_{||}\ (S_z\otimes\sigma_z)$, and potential scattering,
$W\ ({\bf 1}\otimes {\bf 1})$, contributions, where $S^{\pm}$, $\sigma^{\pm}$
are the impurity and the conduction electron spin flip operators, and
$S_z$, $\sigma _z$ the z-components of the impurity spin and the 
conduction electron Pauli matrices, respectively. $J_{\perp}$, $J_{||}$,
$W$ are the respective running coupling constants, with 
$J_{\perp} = J_{||} = 2 W \equiv |V|^2 G_b^{0}(\omega) = |V|^2/|\epsilon_d|$
for $\omega \ll |\epsilon_d|$ for the Anderson model. 
This decomposition leads to the perturbative RG equations for the 
coupling constants within NCA,
\begin{eqnarray}
\frac{dJ_{\perp}} {d{\rm ln} D} &=& - N_0 (J_{||} J_{\perp} +2W J_{\perp})
\label{RG1}\\
\frac{dJ_{||}} {d{\rm ln} D} &=& - N_0 (J_{\perp}^2 + 2 W J_{||})
\label{RG2}\\
\frac{dW} {d{\rm ln} D} &=& - N_0 
(\frac{1}{2} J_{\perp}^2 + \frac{1}{4} J_{||}^2 +W^2) \ .
\label{RG3}
\end{eqnarray}
Eqs.\ (\ref{RG1}--\ref{RG3}) reveal two remarkable facts.
(1) Within NCA the potential scattering  
amplitude $W$ is erroneously renormalized under the RG flow, leading to
a spurious divergence of $W$. 
(2) The RG equations are easily integrated to give ($J_0$ is the
bare coupling),
\begin{eqnarray}
J(D) = \frac{J_0}{1+2N_0 J_0 {\rm ln}\frac{D}{D_0}} \ ,
\end{eqnarray}
i.e. the spin coupling constant $J$ as well as the 
potential scattering amplitude $W$ diverge at the Kondo temperature
$T_K=D_0 {\rm e}^{-1/(2N_0J_0)}$. This demonstrates that for 
$U\to\infty$ the NCA reproduces the Kondo scale correctly, 
however due to an accidential cooperation of spin and potential
scattering terms. 

One may conjecture that this fact, that the NCA does not distinguish between
potential and spin scattering, is the origin why the NCA
gives a qualitatively incorrect description of the Kondo resonance
in a magnetic field, see section \ref{3}: While the divergent 
spin scattering amplitude depends on magnetic field and leads to the 
correct Zeeman splitting of the Kondo resonance \cite{Rosch03}, 
the incorrectly
diverging potential scattering term is insensitive to the magnetic field 
and leads to the spurious third peak produced by the NCA at
$\omega =0$ at finite magnetic field.
\subsection{Perturbative RG for the CTMA}

We now consider the coupling constant renormalization under the
RG flow within CTMA. 
From the analysis of the selfenergies shown in Fig.\ \ref{selfenergyCTMA}
and of the corresponding $G_{d\sigma}$
the CTMA approximation for $\gamma ^{cf}$ is given by the sum of the
contributions shown in Fig.\ \ref{vertices_4p} (1), (2). 
The contribution Fig.\ \ref{vertices_4p} (2) leads to an
additional c--f 
vertex renormalization under cutoff reduction, shown in 
Fig.\ \ref{rgcorrections} (2), 
\begin{eqnarray}
d\Lambda ^{cf, (2)}_{s' \sigma ',s \sigma}
                    = + N_0 \frac{dD}{D} \sum _{s'' \sigma ''}
                    \big[ \Lambda^{(cf)} \big]_{s' \sigma ',s''\sigma ''}
                    \big[ \Lambda^{(cf)} \big]_{s''\sigma '' ,s \sigma} \ ,
\nonumber
\end{eqnarray}
The contribution Fig.\ \ref{vertices_4p} (3) leads to the 
c--b vertex renormalization shown in 
Fig.\ \ref{rgcorrections} (3), which, as a pure potential 
scattering term is not scaling, and, therefore, neglected. 
Adding $d\Lambda ^{cf, (1)}$ and  $d\Lambda ^{cf, (2)}$
and decomposing into $J_{\perp}$, $J_{||}$ and $W$ contributions,
one obtains the RG equations of CTMA,
\begin{eqnarray}
\frac{dJ_{\perp}} {d{\rm ln} D} &=& - 2 N_0 J_{||} J_{\perp}
\label{RGctma1}\\
\frac{dJ_{||}} {d{\rm ln} D} &=& - 2 N_0 J_{\perp}^2 
\label{RGctma2}\\
\frac{dW} {d{\rm ln} D} &=& 0 \ .
\label{RGctma3}
\end{eqnarray}
Inspection of the two terms in Fig.\ \ref{rgcorrections} shows that
they are the direct (2) and the exchange (1) scattering contributions
and, upon collapsing the bare boson lines for 
$\omega \ll |\epsilon _d|$ to the spin coupling,
$-|V|^2G_b^0(\omega ) \to - |V|^2/\epsilon_d = J$, 
that they directly correspond to the two well-known contributions to
the perturbative one-loop $\beta$--function of the Kondo problem 
\cite{Anderson70,Hewson93}.
The failure of the NCA in the weak coupling regime is, thus, traced
back to the fact that it incorporates only the exchange term, and the
direct term is missing. 
This is cured by the CTMA: The RG equations (\ref{RGctma1}--\ref{RGctma3})
are identical to the well-known perturbative RG equations of the 
original Kondo model. 
This proves that the CTMA incorporates the complete Kondo physics
also in the weak and intermediate coupling regime, where the
perturbative RG is valid. 
One may expect that, as a consequence, the CTMA correctly describes the
Zeeman-like splitting of the Kondo resonance in a magnetic field,
since it is quantitatively correct for the susceptibility as well
(see section \ref{6}). 


\begin{thebibliography}{99}
\bibitem{remark} Note, however, that CTMA includes all potential
scattering terms up to $O(\Gamma ^4 )$.
\end{thebibliography}
\end{document}